\documentclass[aps,prl,twocolumn,groupedaddress,showpacs,showkeys]{revtex4}
\usepackage{graphicx}
\usepackage{amssymb}
\usepackage{aas_macros}

\begin{document}
\title{Photon-axion conversion in Active Galactic Nuclei?}
\author{Nicola Bassan}
\email[]{bassan@sissa.it}
\affiliation{SISSA,V. Beirut 2-4, I-34014 Trieste, Italy}
\author{Marco Roncadelli}
\email[]{marco.roncadelli@pv.infn.it}
\affiliation{INFN, Sezione di Pavia, Via A. Bassi 6, I-27100 Pavia, Italy}

\date{\today}

\begin{abstract}
Axion-Like Particles (ALPs) are the focus of intense current research. We analyze photon-ALP conversion in the context of relativistic jet models of 
Active Galactic Nuclei (AGN) for more than 100 sources. Contrary to previous claims, we find that this process cannot occur above $100 \,{\rm GeV}$ regardless of the actual AGN model and the values of ALP parameters. This result rules out a proposed strategy to bypass the cosmic opacity above $100 \,{\rm GeV}$, as apparently required by observations. We also show that for some AGN an observable effect can show up in the $X$ and soft $\gamma$-ray bands. 
\end{abstract}


\maketitle

{\it Introduction} -- Phenomenological as well as conceptual arguments entail that the Standard Model (SM) should be regarded as the low-energy manifestation of a more fundamental unified theory of all interactions. Invariably, a theory of this sort involves a set of new particles. Therefore, the SM lagrangian is expected to be modified by small terms describing interactions among known and new particles. Many extensions of the SM which have attracted considerable interest in the last few years predict the existence of Axion-Like Particles (ALPs). They are spin-zero light bosons defined by the low-energy effective lagrangian
\begin{equation}
\label{a1a}
{\cal L}_{\rm ALP} \ = \ 
\frac{1}{2} \, \partial^{\mu} \, a \, \partial_{\mu} \, a - \frac{1}{2} 
\, m^2 \, a^2 - \frac{1}{4 M} \, F^{\mu \nu} \, \tilde F_{\mu \nu} \, a~,
\end{equation}
where $F^{\mu \nu}$ is the electromagnetic field strength, $\tilde F_{\mu \nu}$ is its dual, $a$ denotes the ALP field whereas $m$ stands for the ALP mass. According to this view, it is assumed $M \gg G_F^{- 1/2} \simeq 250 \, {\rm GeV}$. On the other hand, it is supposed that $m \ll G_F^{- 1/2} 
\simeq 250 \, {\rm GeV}$. Besides than in four-dimensional models~\cite{masso1,masso2,coriano1,coriano2}, ALPs naturally arise in the context of compactified Kaluza-Klein theories~\cite{kk} as well as in superstring theories~\cite{superstring,superstring2}. Moreover, it has been argued that an ALP with mass $m \sim 10^{-33} \, {\rm eV}$ is a good candidate for the quintessential dark energy~\cite{carroll}. The standard Axion~\cite{axion,axion2} is the archetype of ALPs and is characterized by a specific relation between $M$ and $m$, while in the case of {\it generic} ALPs $M$ and $m$ are assumed {\it unrelated}. So, the peculiar feature of ALPs is the trilinear $\gamma$-$\gamma$-$a$ vertex described by the last term in ${\cal L}_{\rm ALP}$, whereby one ALP couples to two photons. 

Because of such a vertex, ALPs can be emitted by various astronomical objects, and this fact yields strong bounds: $M > 0.86 \cdot 10^{10} \, {\rm GeV}$ for $m < 0.02 \, {\rm eV}$~\cite{Zioutas2005} and  $M > 10^{11} \, {\rm GeV}$ for $m < 10^{- 10} \, {\rm eV}$~\cite{Raffelt1990,Raffelt19902,Raffelt19903}. Moreover, the same $\gamma$-$\gamma$-$a$ vertex makes the interaction eigenstates differ from the propagation eigenstates, and so photon-ALP oscillations show up when an external magnetic (electric) field is present~\cite{Sikivie1984,Sikivie19843,Sikivie19844}.

Several attempts have recently addressed the astrophysical implications of ALPs. In this Letter, particular attention will be paid to very-high-energy (VHE) effects, which can be summarized as follows. Below $100 \,{\rm GeV}$ photon propagation throughout cosmic distances is basically unaffected by {\it any} cosmic background radiation~\cite{MarcoPhlB}, whereas above $100 \,{\rm GeV}$ Extragalactic Background Light (EBL) makes the horizon of the observable Universe rapidly shrink as the energy further increases, since the process $\gamma \gamma \to e^+ e^-$ becomes an important source of opacity~\cite{Aharonian00}.  As stressed above, photon-ALP oscillations require the presence of a magnetic field, and in fact specific situations have been envisaged, like a magnetic field in the source, cosmic magnetic fields~\cite{PAuger,dpr} along the line of sight and the Galactic magnetic field. Actually, photon-ALP oscillations have been proposed as a means to circumvent the cosmic opacity -- as apparently required by observations -- by relaying upon the fact that ALPs do not suffer EBL absorption. Two specific scenarios have been put forward:
\begin{itemize}
\item Photon-ALP oscillations are supposed to occur only in intergalactic space~\cite{Marcorev,Marcomon};
\item Photo-to-axion conversion is assumed to take place in the source and to be followed by back-conversion in the Galaxy~\cite{Hooper2,Fairbairn}.
\end{itemize}

Photon-ALP conversion in the source has been envisaged to take place inside the relativistic jets of Active Galactic Nuclei (AGN)~\cite{Hooper,Sigl}. Yet, no extensive quantitative analysis has been carried out so far. Our aim is to fill this gap.  We begin by framing our analysis within leptonic AGN models, but later we will consider hadronic AGN models as well.

{\it Photon-ALP oscillations} -- For our needs, it is sufficient to consider the propagation of a photon beam in a constant magnetic field ${\bf B}$. In such a situation, the probability that a photon converts into an ALP after a distance $x$ can be computed exactly and reads~\cite{Sikivie19844}
\begin{equation}
\label{a2}
P_{\gamma \leftrightarrow a}(x) = \left( \frac{B_T}{M \, {\Delta}_{\rm osc}}  \right)^2 \  {\rm sin}^2
\left( \frac{\Delta_{\rm osc} \, x}{2} \right)~,
\end{equation}
with
\begin{equation}
\label{a3}
{\Delta}_{\rm osc} \equiv 
\left[\left( \frac{7 \alpha}{90 \pi} \frac{B_T^2}{ B_{\rm cr}^2} +
\frac{m^2 - {\omega}_{\rm pl}^2}{2 E} \right)^2 + 
\left( \frac{B_{\rm T}}{M} \right)^2 \right]^{1/2}~, 
\end{equation}
where  
\begin{equation}
{\omega}_{\rm pl}  \equiv  \left( \frac{4 \pi \alpha n_e}{m_e} \right)^{1/2} \simeq 3.69 \cdot 10^{- 11} \, \left( 
\frac{n_e}{{\rm cm}^{- 3}} \right)^{1/2} \, {\rm eV}
\end{equation}
stands for the plasma frequency of the medium, $m_e$ is the electron mass, $n_e$ is the corresponding electron density, $B_T$ is the component of ${\bf B}$ transverse to the beam and $B_{\rm cr} \simeq 4.41 \cdot 10^{13} \, {\rm G}$ is the critical magnetic field.

{\it AGN leptonic models} -- Among the various kinds of AGN, blazars are the ones that look brighter in the VHE band. They are modelled as a pair of  oppositely directed relativistic jets --  one pointing towards the observer -- associated with a rotating supermassive black hole. The jets are supposed to be powered either by magnetic fields twisted by differential rotation of the accretion disk surrounding the black hole~\cite{Meier} or by the energy extracted from the black hole ergosphere~\cite{Blandford77}. Photon emission is assumed to arise from very energetic electrons contained in one or more blobs with typical size of about $10^{16} \, {\rm cm}$ moving along the jet~\cite{CelottiGhisellini07}. In order to explain the strong observed variability of jet emission more than one blob is needed in some cases. 

A crucial ingredient of this picture is the presence of a magnetic field ${\bf B}$ in the jets, which consists of one or more domains larger than the blob~\cite{Naturejet}. As a consequence, the electrons in the blobs radiate through the Synchro Self Compton (SSC) mechanism. Namely, relativistic electrons emit synchrotron radiation up to the X-ray band, whose energy is boosted up to the ${\rm TeV}$ region by Inverse Compton scattering on their parent electrons. It can also happen that low-energy photons coming from outside the blob are upscattered in the same fashion, a situation called External Compton (EC) scattering. This simple model is able to successfully account for the spectra of more than 100 blazars.

\begin{figure}
\includegraphics[width=\columnwidth]{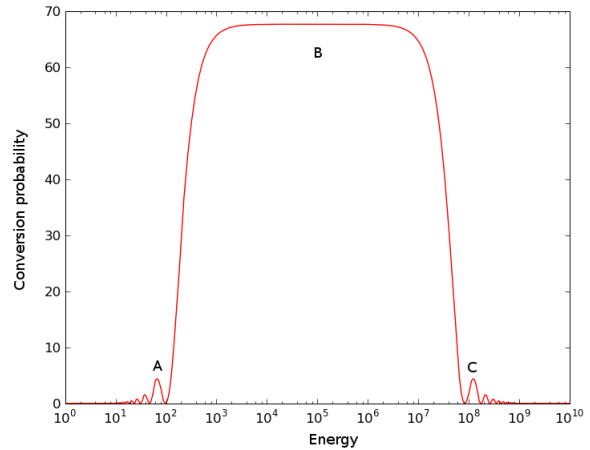}
\caption{\label{fig:prob} We plot $P_{\gamma \leftrightarrow a}$ for $2251+158$ vs. $E$ (eV). 
This figure is obtained for $M=10^{11}\,{\rm GeV}$, $m=10^{-10}\,{\rm eV}$.}
\end{figure}

We begin by considering the emission from a single blob. We use as input parameters in $P_{\gamma \leftrightarrow a}(x)$ the values listed in Table 1 of ref.~\cite{CelottiGhisellini07} and we assume that photon-ALP conversion occurs inside one single blob (this restriction will subsequently be relaxed). We first compute the survival probability $P_{\gamma \leftrightarrow \gamma} \equiv 1 - P_{\gamma \leftrightarrow a}$ for a photon inside a blob and next we convolve it with the pure SSC spectrum, thereby predicting the AGN spectra in the presence of photon-ALP oscillations (we are disregarding for simplicity the existence of extragalactic magnetic fields, which can give a further contribution and should be considered in a more complete analysis~\cite{dpr}). 

\begin{figure}
\includegraphics[width=\columnwidth]{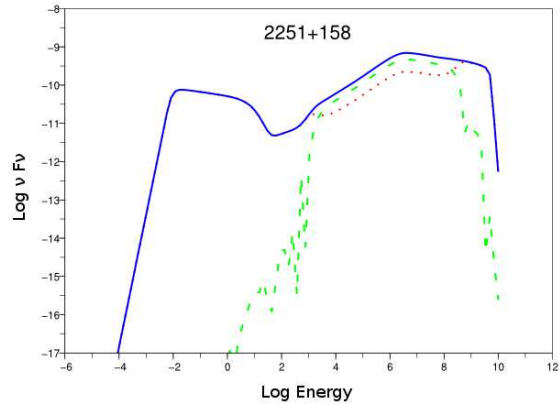}
\caption{\label{fig:spec} We plot $\nu F_\nu$ (${\rm erg \, cm^{-2} \, s^{-1}}$) vs. $E$ (eV) for  $2251+158$. The solid line is the spectrum as computed in ref.~\cite{CelottiGhisellini07} while the dotted line is the spectrum including ALP-Photon oscillation. Dashed line represents the difference between the two. This figure is obtained for $M=10^{11}\,{\rm GeV}$, $m=10^{-10}\,{\rm eV}$.}
\end{figure}

We have analyzed more than $100$ sources, assuming $10^{-10}\, {\rm eV} \leq m \leq 10^{-6} \, {\rm eV}$ and $M=10^{11}\,{\rm GeV}$ as illustrative values. We find $P_{\gamma \leftrightarrow a} > 0.1$ for $12$ of them and $P_{\gamma \leftrightarrow a} > 0.5$ only for 2 sources: $0528+134$ and $2251+158$. Let us focus for definiteness on $2251+158$. We get a non-negligible conversion only within a {\it finite} energy range and the corresponding behaviour of $P_{\gamma \leftrightarrow a}$ is plotted in Fig. (\ref{fig:prob}). Surprisingly, the conversion happens to take place at energies {\it much lower} than the ones previously reported in the literature~\cite{Hooper,Hooper2,Fairbairn,Sigl}. Moreover, while the existence of the low-energy cut-off is commonly appreciated, the relevance of the high-energy one is often neglected in practice~\cite{footQED}. Still, the existence of such a high-energy cut-off can be very important in some astrophyisical settings. Indeed, in the present discussion the above behaviour of $P_{\gamma \leftrightarrow a}$ emerges for {\it all} considered sources, and in no case does a nonvanishing $P_{\gamma \leftrightarrow a}$ show up for $E > 10 \, {\rm GeV}$. 

Remarkably enough, the shape of $P_{\gamma \leftrightarrow a}$ exhibited in Fig. (\ref{fig:prob}) can be understood in an intuitive fashion. To see this, we note that from Eq. (\ref{a2}) it follows that a necessary condition for $P_{\gamma \leftrightarrow a}$ to be maximal is 
\begin{equation}
\label{a3b}
{\Delta}_{\rm osc}  \simeq \frac{B_T}{M}~. 
\end{equation}
In order to work out the relevance of Eq. (\ref{a3b}), it proves instrumental to define
\begin{equation}
\label{a4}
m_* \equiv 10^{-6 }\left| \left( \frac{m}{10^{- 6} \, {\rm eV}} \right)^2 - \left( \frac{{\omega}_{\rm pl}}{10^{- 6} \, {\rm eV}} \right)^2 \right|^{1/2} \, {\rm eV}~,
\end{equation}
so that we have $m_* \simeq m$ for $m \gg \omega_{\rm pl}$ and $m_* \simeq \omega_{\rm pl}$ for $m \ll \omega_{\rm pl}$. In addition, we set
\begin{equation}
\label{a5}
E_* \equiv \frac{|m^2 - {\omega}_{\rm pl}^2 | M}{2 B_T} 
\end{equation}
and
\begin{equation}
\label{a6}
E_{**} \equiv \frac{2}{7} \left(\frac{45 \pi}{\alpha} \right) \left(\frac{B_{\rm cr}}{B_T} \right) 
\left(\frac{B_{\rm cr}}{M} \right)~. 
\end{equation}
Correspondingly, Eq. (\ref{a3}) can be rewritten as
\begin{equation}
\label{a7}
{\Delta}_{\rm osc} =  \left( \frac{B_T}{M} \right)\left[ \left( \frac{E}{E_{**}} + {\epsilon}(m - {\omega}_{\rm pl}) \, \frac{E_*}{E} \right)^2 + 1 \right]^{1/2}
\end{equation}
where ${\epsilon}(m - {\omega}_{\rm pl})$ denotes the sign of its argument. Now, by combining Eqs. (\ref{a3b}) and (\ref{a7}), it is straightforward to 
check that $P_{\gamma \leftrightarrow a}$ is maximal in the so-called {\it strong-mixing regime} defined by $E_* < E < E_{**}$, which is indeed the range $A - C$ shown in Fig. (\ref{fig:prob}). We stress that $E_*$ and $E_{**}$ are {\it independent} of the size of the magnetized region in which photon-ALP takes place, and moreover $E_{**}$ is also {\it independent} of $m$ and ${\omega}_{\rm pl}$: these facts play a key-role throughout the subsequent investigations.

We proceed by extending our analysis to the many-blob case. What happens can be sketched as follows. Focusing one one generic blob, on top of the single-blob behaviour discussed above, the EC emission from blobs closer to the AGN centre has to be taken into account. Furthermore, we use the same assumptions and scaling laws for the blobs as outlined in ref.~\cite{SSC}. We have handled this more complicated situation by means of a numerical code~\cite{footPy}. In this case, $P_{\gamma \leftrightarrow a}$ turns out to be vanishingly small even for $E_* < E < E_{**}$, simply because the decrease of the blob size -- needed to keep the total resulting luminosity constant -- drastically reduces photon-ALP conversion according to Eq. (\ref{a2}).

\begin{figure}
\includegraphics[width=\columnwidth]{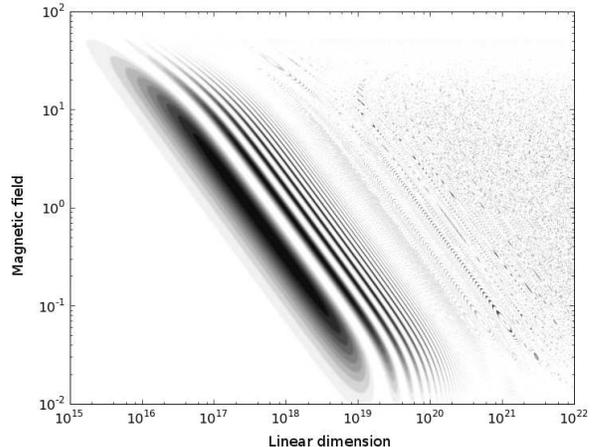}
\caption{\label{fig:param} We plot $P_{\gamma \leftrightarrow a}$ for $E = 10 \,{\rm MeV}$ photons in a $B$ (G) vs. linear dimension (cm) plane. White corresponds to no conversion at all while black corresponds to $P_{\gamma \leftrightarrow a} > 0.9$. This figure is obtained for $M=10^{11}\,{\rm GeV}$, $m=10^{-8}\,{\rm eV}$, $n_{e}=10^3\,{\rm cm^{-3}}$.}
\end{figure}

{\it AGN hadronic models} --  A natural question is whether $P_{\gamma \leftrightarrow a}$ can be sizeable for $E > 100 \, {\rm GeV}$ within AGN hadronic models. Remarkably, this issue can be settled by merely performing an analysis of the ALP parameters space. The result is exhibited in Fig. (\ref{fig:param}). We see that $P_{\gamma \leftrightarrow a}$ is non-vanishing inside a series of elongated sets which move in the parameter space towards wider and less magnetized AGN jet regions as $E$ increases. Specifically, in order to have a sizeable $P_{\gamma \leftrightarrow a}$ for $E > 100 \, {\rm GeV}$, Eq. (\ref{a6}) requires
\begin{equation}
\label{a8}
B_T < 4.5 \cdot 10^{-4} \left(\frac{10^{10} \,{\rm GeV}}{M} \right) {\rm G}~.
\end{equation}
Taking as before $M \simeq 10^{11} \, {\rm GeV}$, we find that $B_T \lesssim 4.5 \cdot 10^{-5} \, {\rm G}$ is needed. Yet, hadronic models demand $10 \, {\rm G} \lesssim B \lesssim  10^2 \, {\rm G}$ inside the jet~\cite{Aharonian00}, thereby implying that $P_{\gamma \leftrightarrow a}$ is {\it always} negligibly small for $E > 100 \, {\rm GeV}$.

\begin{figure}
\includegraphics[width=\columnwidth]{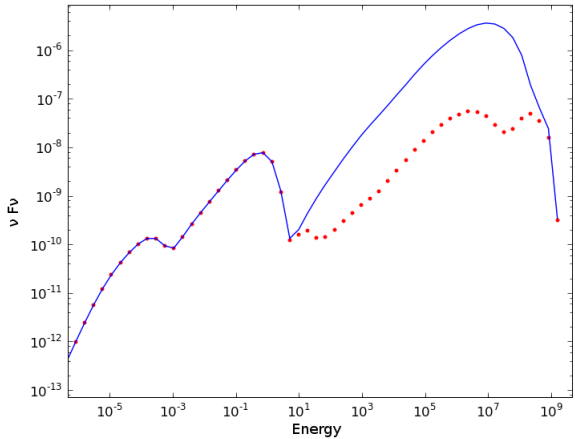}
\caption{\label{fig:multi} We plot  $\nu F_\nu$ (${\rm erg \, cm^{-2} \, s^{-1}}$) vs. $E$ (eV) for a randomly generated source with 10 emitting blobs. The solid line represents the pure SSC spectrum, while the dotted line gives the same spectrum including photon-ALP oscillations.}
\end{figure}

{\it Discussion} --  So far, we have supposed that photon-ALP conversion takes place inside the blobs, but this might not be the case. Hence, we have run the multi-blub simulation for 10 blobs allowing for the conversion to happen also in the intra-blob region of the jet. The result is depicted in Fig. (\ref{fig:multi}). As it is evident, now $P_{\gamma \leftrightarrow a}$ can be sizeable and quantitatively the situation happens to be (just by chance) practically the same as in the single-blob case. It goes without saying that the same argument can be applied to the single-blob case, possibly resulting in an enhanced $P_{\gamma \leftrightarrow a}$~\cite{footelse}.

On the basis of our previous finding, one might think that a magnetic field smaller than $B_T \lesssim 4.5 \cdot 10^{-5} \, {\rm G}$ could give rise to a successful conversion if the model-dependent constraint $10 \, {\rm G} \lesssim B \lesssim  10^2 \, {\rm G}$ were relaxed. However, this turns out to be an impossible task. For, it does not matter to have $E_{**} > 100 \, {\rm GeV}$ if $P_{\gamma \leftrightarrow a}$ is small by its own. As it is clear from Eq.~(\ref{a2}), we also have to require the sine argument to be great enough so as not to kill the conversion probability. It is easy to see that in the strong mixing regime the sine argument is $B x/2 M$, which is of order $1$ provided that ${\bf B}$ is coherent on a scale of about $10^{21} \, {\rm cm}$. Yet, observations~\cite{Naturejet} tell us that at distances greater than $10^{18} \, {\rm cm}$ from the black hole, ${\bf B}$ becomes chaotic and looses its coherence, thereby destroying photon-ALP oscillations. For the same reason, we conclude that no conversion for $E > 100 \,{\rm Gev}$ takes place in the radio lobes opening at the end of the jet.

{\it Conclusions} -- We have analyzed photon-ALP conversion in the context of AGN relativistic jet models. We have found that -- contrary to previous claims~\cite{Hooper,Hooper2,Fairbairn} -- this process cannot occur above $100 \,{\rm GeV}$ regardless of the actual AGN model and the values of ALP parameters. Thus, only photon-ALP oscillations occurring in {\it intergalactic space} can successfully circumvent the cosmic opacity in the VHE band~\cite{Marcorev,Marcomon}. We have also shown that for some sources an observable effect can show up in the $X$ and soft $\gamma$-ray bands~\cite{footlong}. 

{\it Note added} -- Just before the submission of this Letter, we have become aware of a very recent paper~\cite{prada} which discusses photon-ALP conversion in the source {\it and} in intergalactic space. This paper considers only 2 sources: 3C279 and PKS 2155-304 (see their Table 2). In either case, they find that for $E > 100 \, {\rm GeV}$ no conversion inside the source takes place, in agreement with our findings.  So, the problem of the cosmic opacity for these sources can indeed be side-stepped only within the scenario advocated in refs.~\cite{Marcorev,Marcomon}. For $E < 100 \, {\rm GeV}$ things are different, and we agree with ref.~\cite{prada} that a full-fledged analysis should take into account photon-ALP conversion both in the source and in intergalactic space, as stated in the text (such an analysis is currently being performed for the $X$ and soft $\gamma$-ray emission from the sources listed in Table 1 of ref.~\cite{CelottiGhisellini07}).


\begin{thebibliography}{35}
\bibitem{footPy}
The code was written using Python and relying on Numpy and Scipy modules.
\bibitem{footelse}
This point will be addressed in more detail elsewhere.
\bibitem{footlong}
This issue will be the subject of a longer publication.
\end{thebibliography}
\end{document}